\documentclass[pre,preprint,nofootinbib,a4paper,11pt]{revtex4}

\usepackage[centertags]{amsmath}
\usepackage{amsfonts}
\usepackage{amssymb}
\usepackage[sort&compress]{natbib}
\usepackage[hypertex]{hyperref}

\DeclareMathOperator{\Sp}{Sp}

\newcommand{\lan}{\langle}

\newcommand{\ran}{\rangle}

\newcommand{\vf}{\varphi}

\newcommand{\s}{\sigma}

\newcommand{\Si}{\Sigma}

\newcommand{\al}{\alpha}

\newcommand{\be}{\beta}

\newcommand{\ga}{\gamma}

\newcommand{\de}{\delta}

\newcommand{\la}{\lambda}

\newcommand{\ups}{\upsilon}

\begin{document}

\title{Stochastic deformation of a thermodynamic symplectic structure}

\date{\today}

\author{P.O. Kazinski}

\email{kpo@phys.tsu.ru}

\affiliation{Physics Faculty, Tomsk State University, Tomsk, 634050 Russia}

\begin{abstract}

A stochastic deformation of a thermodynamic symplectic structure is studied. The stochastic deformation procedure is analogous to the deformation of an algebra of observables like deformation quantization, but for an imaginary deformation parameter (the Planck constant). Gauge symmetries of thermodynamics and corresponding stochastic mechanics, which describes fluctuations of a thermodynamic system, are revealed and gauge fields are introduced. A physical interpretation to the gauge transformations and gauge fields is given. An application of the formalism to a description of systems with distributed parameters in a local thermodynamic equilibrium is considered.

\end{abstract}

\pacs{05.70.Ln,05.40.-a}


\maketitle

\section{Introduction}

A geometric description of thermodynamic systems has a long history and was initiated by Gibbs \cite{Gibbs}. In this approach, equations of state of a thermodynamic system are represented by a surface in a space of thermodynamic parameters. Later this geometric formalism was developed in the works \cite{Carath,Lande,ErhenA,Tisz,Jauc,Call} on axiomatic foundations of thermodynamics, where its laws were formulated in terms of differential forms. Due to Hermann \cite{Herm} symplectic and contact geometries acquire a distinctive form in thermodynamics, though some elements of these geometries were introduced by Gibbs as well. Notions of symplectic, contact, Riemannian and Finslerian geometries in thermodynamics get a further development in the papers \cite{Weinh,SalIhrBer,MrNuSchSa,Rupp,Bena,BalVal,PreVar,AlQuSa,Vaca}. Since the symplectic structure arose, a strong analogy of thermodynamics with classical mechanics and optics was understood soon afterwards \cite{Peter,Gilm,PoSte,Arn,GrOtt,Sien,MarGam,Raje}. Thermodynamics was realized as a Hamiltonian dynamical system. The next logical step is to ``quantize'' such a dynamical system in order to describe its fluctuations, as it was posed in \cite{Gilm}. At the same time this ``quantization'' is not a quantal one, which is given in \cite{Raje}, but has to result in a Fokker-Planck type equation for a probability distribution of thermodynamic variables. The inverse procedure of ``dequantization'' (the weak noise limit) of the Fokker-Planck equation is of a common knowledge \cite{KuMaKi,vKam,GrRoTe} and also endows thermodynamics with the symplectic structure. In this paper, we shall see that the desired ``quantization'' is a stochastic deformation procedure proposed in \cite{sd}, which is an analog of the algebraic approach to quantization known as deformation quantization \cite{BerezQ,BFFLS,Fed,Kont}.

A theory of fluctuations of thermodynamic quantities is well-known both for equilibrium \cite{Eins} and non-equilibrium processes, and has a huge literature (see, e.g., \cite{Ons,NicPrig,dGroMaz,OnsMach,vKam,KuToHa,LandLifstatI,Grah,BochKuz,Strat,ZuMoRo,JCVL,Berd}). Thus the aim of ``quantization'' lies in a construction of stochastic mechanics\footnote{To avoid misunderstanding we point out that we use the term ``stochastic mechanics'' in a general sense like a notion that unifies various physical models with dynamics obeying some kind of the master equation. It is not Nelson's Stochastic mechanics \cite{Nel}.} by means of a procedure similar to an ordinary quantization, while it should give rise to the standard theory of fluctuations and be equivalent to operator methods of stochastic mechanics \cite{Nam,Par,Ris,ZJ,Vasil}. Stochastic deformation applied to thermodynamics complies with these requirements and reproduces known operator methods in some particular gauges. Gauge transformations and gauge fields are necessary ingredients of stochastic deformation. Furthermore, as we shall see, they are intrinsic to thermodynamics. Transformations of this kind arise occasionally in the literature \cite{GrRoTe,SRPHH} as some tricks to prove, for example, an equivalence of the Doi \cite{Schon,Doi,MatGlas,MSST} and Martin-Siggia-Rose \cite{Nam,MSR,Phyth,Jans,DomPel,GLSM,AdVaPi,JorSukhPil} formalisms \cite{LefBir}, or to establish certain properties of operator's spectrum of the master equation, or they appear in the form of nonstandard inner products \cite{KuToHa,vKam,Kaw}, or as the $\al$ degree of freedom in Umezawa's thermo field dynamics \cite{Umez} and so on. As long as thermodynamics and stochastic mechanics have such symmetries, it is of importance to represent their governing equations in an explicitly invariant form like, for example, to represent the Maxwell equations in an explicitly Lorentz and gauge covariant fashion.

Stochastic deformation of the thermodynamic symplectic structure reveals, firstly, a perfect analogy of thermodynamics and stochastic mechanics on the one hand with classical mechanics and its quantum deformation on the other hand. Secondly, it allows us to discover natural gauge structures of thermodynamics and stochastic mechanics and formulate them in gauge invariant forms.  In comparison with the standard operator approach to stochastic mechanics, we effectively introduce an additional St\"{u}ckelberg field and corresponding gauge fields, which result in gauge invariant dynamics. A significant feature of these rather formal manipulations is that the St\"{u}ckelberg field possesses a physical interpretation. It is an entropy of the thermodynamic system, while the gauge fields are external thermodynamic forces acting on it.

It is worthy to mention the works that are closely related to the subject matter of the paper. A probabilistic stochastic deformation of a linear symplectic structure was studied in the papers \cite{Zambr1} within the framework of Zambrini's Euclidean quantum mechanics \cite{Zambr,BlGaOlk}. The latter stems from Schr\"{o}dinger's works \cite{Schro} made in attempt to give a stochastic interpretation to quantum mechanics. In the papers \cite{Garb}, Lagrangian and Hamiltonian descriptions of Smoluchowski diffusion processes are given as a special case of the general formalism based on diffusion-type equations. Notice also that something similar to the gauge fields we are about to consider arose in \cite{Grah}. Gauge transformations, but of the different type, was introduced into thermodynamics in \cite{BalVal}.

The paper is organized as follows. We start with a formulation of thermodynamics in terms of symplectic geometry (Sec. \ref{thermod}). Then we realize non-equilibrium thermodynamics as a certain Hamiltonian dynamical system and establish its gauge invariance. Here we also introduce gauge fields and provide their physical interpretation. In Sec. \ref{fluct} we consider a stochastic deformation of the obtained Hamiltonian dynamical system and investigate some its properties. We introduce an operator of dissipation, which is the stochastic analog of a dissipation function \cite{OnsMach,LandLifstatI,Rayl}, and express an entropy production through it. As an example we apply the formalism to a thermodynamic system being in a local thermodynamic equilibrium. In conclusion, we outline prospects for further generalizations and research. We assume Einstein's summation rule unless otherwise stated. Latin indices run from $1$ to $d$, where $d$ is a number of extensive variables, and zero index corresponds to time.

\section{Thermodynamics}\label{thermod}

All the statistical properties of a quantum system in a thermostat are determined by the partition function
\begin{equation}\label{stat sum}
    Z(p_1,\ldots,p_{d-1},x^d):=\Sp\exp\left(-\sum\limits_{i=1}^{d-1}p_i\hat{x}^i\right),
\end{equation}
where $\hat{x}^i$ are quantum operators of additive integrals of motion, $x^d$ is some fixed extensive variable, for instance, the volume, and $p_i$ are intensive parameters thermodynamically conjugate to $\hat{x}^i$ or thermodynamic forces. The Hamiltonian and the reciprocal temperature are among these conjugate pairs. Differentiating the partition function with respect to $p_i$ and making the Legendre transform we arrive at the first law of thermodynamics
\begin{equation}\label{1law}
  dS(x)=p_idx^i,
\end{equation}
where $x^i$ are averages of the corresponding operators, $S(x)$ is the entropy of the system, which is the Legendre transform of the Massieu function \cite{Call,Bali,ZuMoRo} $\Phi:=-\ln Z$, and we introduce the intensive parameter $p_d:=\partial_dS(x)$ conjugate to $x^d$. In geometric terms, the first law \eqref{1law} endows the space of states $(x,p)$ of a thermodynamic system with a symplectic structure specified by the symplectic potential $\theta:=p_idx^i-dS$. According to \eqref{1law} the system is confined to the Lagrangian surface of the symplectic $2$-form $d\theta$. Inversely, any Lagrangian surface uniquely projectable to the space of extensive variables $\{x^i\}$ can be locally represented by the equality of the form \eqref{1law} (see, e.g., \cite{Arn}). For reasonable physical systems the Lagrangian surface is uniquely projectible, otherwise a change of thermodynamic forces does not vary extensive parameters, that is, we have a system with zero (generalized) compressibility.

In nonequilibrium with the thermostat, the thermodynamic system moves along the Lagrangian surface. Besides, if we directly (not by means of intensive parameters) change the entropy of the system, the Lagrangian surface \eqref{1law} also evolves. A natural generalization of the first law to nonequilibrium processes looks like
\begin{equation}\label{1law modified}
    dS(t,x)=p_idx^i-H(x,p,t)dt,
\end{equation}
where $H(x,p,t)$ is a thermodynamic force conjugate to time or the Hamilton function\footnote{Of course, this Hamilton function is not related, at least immediately, to the Hamiltonian entering the partition function \eqref{stat sum}. In \cite{Strat,BochKuz} this thermodynamic force is called the kinetic potential.}. The Hamilton function taken on the Lagrangian surface is a source of the entropy and it vanishes if the functional form of the entropy does not change. The non-stationary first law \eqref{1law modified} represents the Hamilton-Jacobi equation. It is valid when it is reasonable to attribute certain values of intensive parameters to the whole thermodynamic system.

The requirement that the system should be confined to the Lagrangian surface restricts the form of the Hamilton function. Here we consider two equivalent mechanisms how to keep the system on the Lagrangian surface
\begin{equation}\label{lagr surf}
    T_i:=p_i-\partial_iS=0.
\end{equation}
The first one is a ``rigid'' method. We demand that $T_i$ are integrals of motion of the Hamilton system associated with \eqref{1law modified}. A general form of a regular in momenta Hamilton function meeting this requirement is
\begin{equation}\label{hamilton func}
    H=c(t)-\partial_tS+T_i\ups^i(t,x)+\frac12T_ig^{ij}(t,x)T_j+\ldots,
\end{equation}
where $\ups^i$ and $g^{ij}=g^{ji}$ are some contravariant tensors, dots denote the terms of a higher order in $T_i$, and $c(t)$ is some function.  Redefining the entropy we eliminate the latter function. Then for a stationary entropy the Hamilton function \eqref{hamilton func} is zero on the Lagrangian surface. The Hamilton equations of motion of the thermodynamic system become
\begin{equation}\label{ham eqs}
    \dot{x}^i=\{x^i,H\}\approx\ups^i,\qquad\dot{T}_i=\partial_tT_i+\{T_i,H\}\approx0,
\end{equation}
where curly brackets denote the Poisson brackets, and approximate equalities mean that we take equations on the Lagrangian surface \eqref{lagr surf}. Now we see that the vector field $\ups^i$ describes a drift of the system. The tensor $g^{ij}$ and higher terms in the expansion \eqref{hamilton func} gain a physical meaning only upon stochastic deformation of the Hamiltonian system, which will be considered below. They are responsible for a probability distribution law of fluctuations of statistical averages.

The Hamiltonian action functional associated with \eqref{1law modified} is obvious. If the expansion of the Hamilton function \eqref{hamilton func} terminates at the term quadratic in $T_i$ and the tensor $g^{ij}$ is nondegenerate, it is not difficult to obtain a Lagrangian form of this action
\begin{equation}\label{action OnsMach}
    S[x(t)]=\int dt\left[\frac12g_{ij}(\dot{x}^i-\ups^i)(\dot{x}^j-\ups^j)+\frac{dS}{dt}\right],
\end{equation}
where $g_{ij}$ is the inverse of $g^{ij}$. This functional is known as the Onsager-Machlup action \cite{OnsMach,TakWat,MarSel}. It measures an entropy change along a trajectory $x(t)$. The first term measures an additional entropy production caused by fluctuations deviating intensive parameters of the system from the Lagrangian surface. This term vanishes in the thermodynamic limit \eqref{ham eqs}. The second term is a change of the thermodynamic entropy. The principle of least action corresponding to \eqref{action OnsMach} says that the system moves to equilibrium with a minimum of the entropy production. Therefore the first term has to be nonnegative for an arbitrary trajectory and the tensor $g^{ij}$ must be positive definite. Further, if the fluctuations are described by a general Markov process, it can be proven \cite{Strat} that $\partial^{ij}_pH$ should be positive definite.

Under the above restrictions on $g^{ij}$ any drift vector field can be represented in the form
\begin{equation}\label{drift}
    \ups^i=g^{ij}(\partial_jS-A_j),
\end{equation}
where $A_i(t,x)$ is some local $1$-form, which we shall call the gauge field. The quantity in the brackets is the total thermodynamic force, while the gauge field $A_i$ is the external force exerting on the system. In particular, the Onsager principle \cite{Ons} postulates the drift \eqref{drift} and that the thermostat acts on the nonequilibrium system as
\begin{equation}\label{ons prpl}
    A_i(t,x)=\partial_iS(t,x_0),
\end{equation}
where $x^i_0$ is the solution of \eqref{lagr surf} at fixed intensive parameters $p_i$ characterizing the thermostat. To provide a stability of the state $x_0$ of the system we have to require a negative definiteness of the Hessian $\partial_{ij}S(x_0)$, otherwise the system is in a phase transition state. If the total thermodynamic force is given,  the relation \eqref{drift} can be only a linear approximation to a nonlinear response of the system to the uncompensated force. Besides, there are systems that do not obey the Onsager principle (see, for physical examples, \cite{NicPrig,Vasil,vKam,Ebel,DomPel,TChChP}).

The first law \eqref{1law modified} with the Hamilton function \eqref{hamilton func} is invariant under the following gauge transformations
\begin{equation}\label{gauge trans}
    p_i\rightarrow p_i+\partial_i\xi(t,x),\qquad
     A_i\rightarrow A_i+\partial_i\xi(t,x),\qquad
     S(t,x)\rightarrow S(t,x)+\xi(t,x).
\end{equation}
Their existence reflects the fact that a gradient part of the external force can be attributed to the system itself redefining its entropy. In other words, these transformations relate equivalent thermodynamic systems (system)+(thermostat). They are not distinguishable within thermodynamics since the total forces do not change under the gauge transformations \eqref{gauge trans}.

Now we are in position to introduce the second method to confine the system to the Lagrangian surface. This method is, of course, equivalent to the first one and based on using an auxiliary compensating field. We postulate that the first law \eqref{1law modified} is invariant under the gauge transformations \eqref{gauge trans}. In addition, we seek the Hamilton function, which does not depend on the entropy. Then a quadratic in momenta Hamilton function\footnote{A generalization of these considerations to Hamilton functions of an arbitrary order in momenta is straightforward, but has no such a suggestive mechanical analogy. Quadratic in momenta Hamilton functions correspond to a Gaussian distribution law of fluctuations $\de$-correlated in time.} providing gauge invariance to \eqref{1law modified} can be cast into the form of the Hamilton function of a charged particle in an electromagnetic field
\begin{equation}\label{hamilton func 1}
    H=\frac12\mathcal{P}_ig^{ij}\mathcal{P}_j-A_0=:K(p,x)-A_0,\qquad\mathcal{P}_i:=p_i-A_i,
\end{equation}
where $A_0(t,x)$ is the auxiliary field transforming under the gauge transformations as
\begin{equation}
    A_0\rightarrow A_0+\partial_t\xi(t,x).
\end{equation}
It keeps an invariance of \eqref{1law modified} with respect to nonstationary gauge transformations. To confine the system to the Lagrangian surface determined by a given  entropy function $S(t,x)$, we have to choose the compensating field $A_0$ so that the Hamilton-Jacobi equation is fulfilled:
\begin{equation}\label{HJE}
    A_0-\partial_tS=\frac12g^{ij}(\partial_iS-A_i)(\partial_jS-A_j).
\end{equation}
If we substitute $A_0$ from the Hamilton-Jacobi equation to the Hamilton function \eqref{hamilton func 1}, we revert to \eqref{hamilton func} with the drift vector field \eqref{drift} establishing the equivalence of two approaches. The gauge invariant total force $A_0-\partial_tS$ is the kinetic part $K$ of the Hamilton function taken on the Lagrangian surface. These two functions, $A_0-\partial_tS$ and $K$, are different representations of the dissipative function $\Psi(X,X)$ introduced in \cite{OnsMach}. The entropy production $\dot{\Si}$ in the whole system (thermostat)+(system) is standardly expressed in terms of the dissipation function
\begin{equation}\label{entropy prod}
    \dot{\Si}\approx\mathcal{P}_i\frac{\partial K}{\partial p_i}=2K,
\end{equation}
where we assume that the rate of an entropy change of the thermostat is $-\ups^iA_i$.

Given $K$ and $A_\mu$ completely define the system and, in particular, its thermodynamic entropy through the Hamilton-Jacobi equation. From mechanical viewpoint a thermodynamic system \eqref{hamilton func 1} tends to an unstable equilibrium of the potential $-A_0$. That is, for any given $x$ we take such an initial momentum $p$ that the ``particle'' hits precisely the equilibrium point with zero velocity. Although this point is unstable equilibrium, it is of course an attractor on the Lagrangian surface. Not any mechanical Hamilton function \eqref{hamilton func 1} having a nonstable equilibrium point in its potential can serve as the Hamilton function for some thermodynamic system even in a neighbourhood of this point. The ``magnetic'' field can freeze the particle (like in a magnetized plasma) so it never reaches the equilibrium. Linearizing particle's equations of motion in a small vicinity of the stationary point $x_0$ and assuming that the Hamilton function is stationary and the fields entering it are smooth enough, one finds that for the isotropic potential $\partial_{ij}A_0(x_0)=\xi g_{ij}(x_0)$ the system never hits the point $x_0$ starting in an arbitrary point of its small vicinity only if
\begin{equation}\label{freezing}
    |\la_i|>4\xi,
\end{equation}
for some $i$, where $\la_i$ are the characteristic numbers
\begin{equation}
    \det(F_{ik}g^{kl}F_{lj}-\la g_{ij})=0,\qquad F_{ij}:=\partial_{[i}A_{j]}(x_0).
\end{equation}
In other words, the condition \eqref{freezing} reminds that the gauge fields $A_i$ should enter the potential $A_0$ according to the Hamilton-Jacobi equation \eqref{HJE}.

\section{Fluctuations}\label{fluct}

Now we turn to fluctuations of statistical averages. As it was shown in \cite{sd}, these fluctuations can be obtained by stochastic deformation of the corresponding Poisson structure. In our case, we shall deform a canonical symplectic structure associated with the thermodynamic system.

Let us briefly recall some basic features of an algebraic stochastic deformation. For more details, an interested reader can consult the paper \cite{sd} and the classical works on deformation quantization \cite{BerezQ,Kont,Fed,BFFLS}. A commutative  associative algebra of classical observables is constituted by real smooth functions over the symplectic space. We deform this algebra in a manner of deformation quantization, but with an imaginary deformation parameter, such that
\begin{equation}
    [\hat{x}^i,\hat{p}_j]=\nu\de^i_j,
\end{equation}
where $\nu$ is the real positive deformation parameter. Hats signify elements of the deformed associative algebra and we imply the Weyl-Moyal star product \cite{BerezMSQ}
\begin{equation}
    \hat{F}\hat{G}=\sum\limits_{n=0}^\infty\frac1{n!}\left(\frac{\nu}2\right)^n\omega^{a_1b_1}\ldots\omega^{a_nb_n}\partial_{a_1\ldots
    a_n}F(z)\partial_{b_1\ldots b_n}G(z),
\end{equation}
where $z\equiv(x,p)$, $a_n,b_n=\overline{1,2d}$, the functions $F(z)$ and $G(z)$
are the Weyl symbols of the corresponding elements of the deformed
algebra, $\omega^{ab}$ is the inverse to the symplectic $2$-form
$\omega_{ab}$. The generators $\hat{x}^i$ and $\hat{p}_j$ of the Heisenberg-Weyl algebra correspond to extensive and intensive parameters of the thermodynamic system. The deformation parameter $\nu$ is not the Planck constant. It characterizes a variance of fluctuations and, as we shall see, is equal to doubled the Boltzmann constant, $\nu=2k_B$, for thermal fluctuations.

Another necessary ingredient of the deformation procedure is the trace functional $\Sp$, which is a linear functional on the deformed algebra mapping to real numbers and vanishing on commutators. An explicit formula for the trace of an element $\hat{F}$ has the form
\begin{equation}
    \Sp\hat F=\int\frac{d^dxd^dp}{(2\pi\nu)^d}F(x,ip).
\end{equation}
A state of a stochastic system is characterized by an element $\hat\rho$ with a unit trace
\begin{equation}
    \Sp\hat\rho=1.
\end{equation}
The pure state is specified by an additional idempotency requirement
\begin{equation}
    \hat\rho^2=\hat\rho.
\end{equation}
An average of some observable $\hat{F}$ over the state $\hat\rho$ is defined by the standard formula
\begin{equation}
    \lan\hat{F}\ran:=\Sp(\hat\rho\hat{F}).
\end{equation}
Thus, for a correct probabilistic interpretation the state $\hat\rho$ should satisfy
\begin{equation}
    \lan\de^d(\hat{x}^i-x^i)\ran\geq0,\quad\forall x\in\mathbb{R}^d.
\end{equation}
The dynamics of the stochastic system in the state $\hat\rho$ are
generated by the element $\hat{H}$ of the deformed algebra, which corresponds to
the Hamilton function $H(t,x,p)$, and obey the von Neumann equation
\begin{equation}
    \nu\dot{\hat\rho}=[\hat H,\hat\rho].
\end{equation}

In the case of a linear symplectic space it is useful to realize the Heisenberg-Weyl algebra as operators acting in the linear space $V$ of smooth real functions on the configuration space. Then, in Dirac's notations, the pure state is represented by\footnote{Similar projectors also arise when describing projected processes \cite{KuToHa}.}
\begin{equation}
    \hat{\rho}=|\psi\ran\lan\vf|,\quad\lan\vf|\psi\ran=1,\qquad|\psi\ran\in V,\quad\lan\vf|\in V^*,
\end{equation}
where the standard inner product is understood. So, the pure state is specified by two real functions on the configuration space. For such states the von Neumann equation is equivalent to two Schr\"{o}dinger-Zambrini (SZ) equations \cite{Zambr,Schro}
\begin{equation}\label{schr eqs}
    \nu\partial_t|\psi\ran=\hat H|\psi\ran,\qquad\nu\partial_t\lan\vf|=-\lan\vf|\hat H.
\end{equation}
Hereinafter, for simplicity, we restrict ourself to the case of at most quadratic in momenta Hamiltonians. Besides, we take the metric tensor $g_{ij}$ to be a constant matrix.

After introducing the stochastic phase $\tilde S(t,x)$ and the probability density function $\rho(t,x)$ to find a system with the values of extensive parameters $x^i$
\begin{equation}\label{repres}
     \psi=\rho\exp(-\nu^{-1}\tilde{S}),\qquad \vf=\exp(\nu^{-1}\tilde{S}),
\end{equation}
the operators of the total forces $\hat{\mathcal{P}}_\mu=\hat p_\mu-A_\mu$ have the averages
\begin{equation}
    \lan\hat{\mathcal{P}}_\mu\ran=\int d^dx \vf(t,x)[-\nu\partial_\mu-A_\mu(t,x)]\psi(t,x)=\lan\partial_\mu\tilde S-A_\mu\ran,\quad\mu=\overline{0,d},
\end{equation}
where, for $\mu=0$, we have used the equations of motion \eqref{schr eqs}. Matching thermodynamics with its deformation, we should identify the phase $\tilde S$ with the thermodynamic entropy $S$. Then the SZ equations \eqref{schr eqs} are invariant with respect to the gauge transformations \eqref{gauge trans} both with the Hamiltonian \eqref{hamilton func} and \eqref{hamilton func 1}.

Consider stochastic deformation of the Hamiltonian dynamical system \eqref{hamilton func}. With the above mentioned identification, one of the SZ equations \eqref{schr eqs} is identically satisfied
\begin{equation}
    \nu\partial_t\lan\vf|=-\lan\vf|(-\partial_tS+\hat{T}_i\ups^i(t,\hat{x})+\frac12\hat{T}_ig^{ij}\hat{T}_j),
\end{equation}
and the other equation becomes the Fokker-Planck equation describing fluctuations of the thermodynamic system
\begin{equation}\label{FPE}
    \partial_t\rho=-\partial_i(-\frac\nu2g^{ij}\partial_j\rho+\ups^i\rho).
\end{equation}
We see that the coefficients of expansion \eqref{hamilton func} in terms of $T_i$ are nothing but the cumulants of the probability distribution of fluctuations.

Deforming the Hamiltonian dynamics generated by \eqref{hamilton func 1} we arrive at two equations: the Fokker-Planck equation \eqref{FPE}, and the Hamilton-Jacobi equation \eqref{HJE} with stochastic correction or the Burgers equation \cite{BlGaOlk},
\begin{equation}
    A_0-\partial_tS=\frac12g^{ij}(\partial_i
    S-A_i)(\partial_j
    S-A_j)+\frac\nu2\partial_i[g^{ij}(\partial_j
    S-A_j)],
\end{equation}
defining the dissipation function $A_0-\partial_tS$. The form of gauge transformations \eqref{gauge trans} and the representation \eqref{repres} show that the gauge group is the Abelian one dimensional Lie group isomorphic to $SO(1,1)$. The SZ equations can be rewritten in an explicitly covariant form with respect to the gauge transformations if we group the two functions $\psi(x)$ and $\vf(x)$ into one vector $\Psi^\al(x)$ and define the pseudo-Euclidean scalar product of such vectors as
\begin{equation}\label{scal prod}
    \frac12\int d^dx\Psi'^\al(x)\eta_{\al\be}\Psi^\be(x)=\frac12\int d^dx[\psi'(x),\vf'(x)]\left[%
\begin{array}{cc}
  0 & 1 \\
  1 & 0 \\
\end{array}%
\right]\left[%
\begin{array}{c}
  \psi(x) \\
  \vf(x) \\
\end{array}%
\right].
\end{equation}
Then the SZ equations \eqref{schr eqs} look like the matrix Schr\"{o}dinger equation
\begin{equation}
    \nu\partial_t\Psi=\left[%
\begin{array}{cc}
  \hat{H} & 0 \\
  0 & -\hat{H}^+ \\
\end{array}%
\right]\Psi,
\end{equation}
where the cross denotes a conjugation with respect to the standard inner product. The evolution is generated by the matrix Hamiltonian, which is skew-adjoint with respect to the scalar product \eqref{scal prod}. On introducing the self-adjoint covariant derivatives
\begin{equation}
    \mathcal{P}^\al_{\mu\be}:=-\omega^\al_\ga(\nu\partial_\mu\de^\ga_\be+\omega^\ga_\be A_\mu),
\end{equation}
where
\begin{equation}
    \omega^\al_\be=\left[%
\begin{array}{cc}
  1 & 0 \\
  0 & -1 \\
\end{array}%
\right],\qquad\omega^2=1,\quad\omega^T\eta=-\eta\omega,
\end{equation}
the SZ equations with the Hamiltonian \eqref{hamilton func 1} read as
\begin{equation}\label{schr eqs matrix form}
    -\mathcal{P}_{0}\Psi=\frac12\mathcal{P}_{i}g^{ij}\mathcal{P}_j\Psi.
\end{equation}
In this form, the SZ equations are very similar to the quantum mechanical Schr\"{o}dinger equation. The difference is that the generator $\omega$ of the Lie algebra $so(1,1)$ appears instead of the generator of $u(1)\simeq so(2)$. Further, the equations of motion \eqref{schr eqs matrix form} are immediately generalized to a non-Abelian case. If we have $N$ identical thermodynamic systems in a thermostat then the global symmetry group for a whole system will be $SO(N,N)\cap Sp(N)$. The symplectic group $Sp(N)$ arises since in the case of $N$ identical systems the matrix $\eta\omega$ is the unit symplectic matrix, which must be preserved by the symmetry transformations. A detailed investigation of peculiarities of non-Abelian systems and their physical interpretation will be given elsewhere. Here we just notice that the group $SU(1,1)$ locally isomorphic to $SO(2,2)\cap Sp(2)/SO(1,1)$ was studied in the context of generalized coherent states \cite{Rase,Perel,Luo,WaSaPa}. The pair $(\eta,\omega)$ is the analog of an almost generalized product structure on the Whitney sum $TM\oplus T^*M$ (see, for example, \cite{Zabz}).

Let us consider how some standard thermodynamic relations look in our framework.
The condition of a detailed balance in some state $|\psi\ran\lan\vf|$ looks like
\begin{equation}\label{det bal}
    \vf(x)\circ\hat{H}\circ\psi(x)=\psi(x)\circ\hat{H}^+\circ\vf(x),
\end{equation}
where $\circ$ means a composition of operators. That is, we can make the gauge transformation so that $\vf=\psi$, the Hamiltonian $\hat{H}=\hat{H}^+$ having a nonpositive spectrum (see, e.g., \cite{Haken,ZJ,vKam,KuToHa}). Then the stochastic phase is identified with a half of the entropy of the whole system (thermostat)+(system). For Hamiltonians of the form \eqref{hamilton func 1} with nondegnerate metric $g^{ij}$ the probability density function $\rho(x)$ in this state\footnote{The use of this probability density function in the study of fluctuations of the Van der Waals gases can be found, for example, in \cite{BeBuTr,AguMal}.} is
proportional to $\exp[2\nu^{-1}\Si(x)]$, where $\Si(x):=S(x)-p_ix^i$, provided the Onsager principle \eqref{ons prpl} is fulfilled. Fixing the extensive variable $x^d$ and applying the WKB-method, we obtain that the leading in $\nu$ terms of the characteristic function $\ln z(p,x^d)$ of the probability distribution $\rho$ at fixed $x^d$ take the form
\begin{equation}\label{char func}
    \ln z(p,x^d)=2\nu^{-1}\ln Z(p,x^d)+\Sp\ln|\partial^{ij}_p\Phi(p,x^d)|/2+\ldots
\end{equation}
Thus we arrive at the well-known result that in the leading order the correlators of statistical averages are proportional to the correlators computed with the help of the partition function \eqref{stat sum}.

In the Heisenberg representation, the operator of the system entropy change is
\begin{equation}
    \dot{\hat{S}}=\partial_t\hat{S}+\nu^{-1}[\hat{S},\hat{H}]=\partial_t\hat{S}+\partial_i\hat{S}\frac{\partial\hat{K}}{\partial\hat{p}_i},
\end{equation}
where $\partial_i\hat{S}\partial/\partial\hat{p}_i$ is a differentiation of the Heisenberg-Weyl algebra acting on its generators in an obvious manner. Consequently, the gauge invariant entropy production in the whole system (thermostat)+(system) is given by
\begin{equation}\label{entropy prod st}
    \dot{\hat\Si}:=(\partial_i\hat{S}-\hat{A}_i)\frac{\partial\hat{K}}{\partial\hat{p}_i},
\end{equation}
where the dot is just a notation in the case of $F_{ij}\neq0$. The relation \eqref{entropy prod st} is the stochastic (noncommutative) analog of its thermodynamic counterpart \eqref{entropy prod}. It is reasonable to define the operator of a purely fluctuational entropy production as
\begin{equation}
    -\dot{\hat{x}}^i\hat{T}_i=-\nu^{-1}[\hat{x}^i,\hat{H}]\hat{T}_i,
\end{equation}
that follows from a path-integral representation of its average. This kind of the entropy production disappears in the thermodynamic limit. If the system with the Hamiltonian \eqref{hamilton func 1} is in the state, where the detailed balance takes place, the average entropy production $\dot{\hat{\Si}}$ is zero. The average of the fluctuational term becomes
\begin{equation}\label{entropy prod st micr}
    -\nu^{-1}\lan[\hat{x}^i,\hat{K}]\hat{T}_i\ran=-\nu\lan g^{ij}\partial_{ij}\Si\ran=2\lan g^{ij}\partial_i\Si\partial_j\Si\ran.
\end{equation}
It is positive and of the order of $\nu$. This is valid in the leading order in $\nu$ for any Hamiltonian that allows a detailed balance. Thus the entropy production $\dot{\hat{\Si}}$ is caused by equalizing large (macroscopic) differences of intensive parameters of the system and the thermostat, whereas the fluctuational term is responsible for the entropy change made by equalizing small (microscopic) differences of intensive parameters originating from fluctuations. The microscopic deviations can be estimated from the well-known thermodynamic uncertainty relation (see, e.g., \cite{Rupp}). In the state with a detailed balance, we have (for a proof see, e.g., \cite{sd})
\begin{equation}
    \lan(x^i)^2\ran\lan(\partial_i\Si)^2\ran\geq\nu^2/4,\qquad \text{(no summation)}.
\end{equation}

A natural generalization of the above considerations to more realistic nonequilibrium systems possessing spatial gradients of intensive parameters is straightforward. We break the system to subsystems of a fixed volume, which are small enough to have homogeneous intensive parameters and sufficiently large to apply a thermodynamic description to them, i.e., the system is in a local thermodynamic equilibrium. Then the simplest model following from first principles of thermodynamics \eqref{1law modified} and \eqref{drift} prescribes a diffusion-like evolution \cite{Ons,Call,BeMoBa,ZuMoRo,BalVal}
\begin{equation}\label{drift field model}
    \partial_t\phi^a(t,x)=\int dy g^{ab}(x,y)\left(\frac{\de S[\phi]}{\de\phi^b(t,x)}-\frac{\de S[\phi]}{\de\phi^b(t,y)}\right),\qquad g^{ab}(x,y)=g^{ab}(y,x)=g^{ba}(x,y),
\end{equation}
where $\phi^a(t,x)$ are densities of the extensive variables except the volume, the functional $S[\phi]$ is the thermodynamic entropy of the whole system, and $g^{ab}$ is some positive definite matrix for any $x$ and $y$, which measures a linear response of the extensive variable $a$ of the subsystem in the point $x$ on a difference of thermodynamic forces $b$ of the subsystems located at $x$ and $y$. The use of the linear response relation is justified by small, by construction, deviations of intensive parameters of neighboring subsystems. The total values of extensive variables are conserved by the evolution \eqref{drift field model}. The Hamilton functional of the form \eqref{hamilton func 1} corresponding to Gaussian fluctuations becomes
\begin{equation}\label{hamilton func 1 field}
    H[t,\phi,\pi]=\frac14\int dxdy(\pi_a(x)-\pi_a(y))g^{ab}(x,y)(\pi_b(x)-\pi_b(y))-A_0[t,\phi],
\end{equation}
where $\pi_a$ are the intensive parameters canonically conjugate to $\phi^a$. If external thermodynamic force fields are applied, the momenta should be replaced by the covariant derivatives \eqref{hamilton func 1}. The Hamiltonian formalism is also preferred to the Lagrangian one as the Onsager-Machlup action \eqref{action OnsMach} is nonlocal for local $g^{ab}$. The functional of a thermodynamic entropy increases with the evolution \eqref{drift field model} and acquires maximum when all the intensive parameters become homogeneous.

Upon stochastic deformation of the model \eqref{hamilton func 1 field} we see that the fluctuating system possesses the states in which the detailed balance \eqref{det bal} takes place:
\begin{equation}
  \psi/\vf\propto\exp\int\la_a\phi^a(x)dx,
\end{equation}
where $\la_a$ are some constants and the stochastic phase $\nu\ln\vf$ is the thermodynamic entropy. We divide the whole system into a thermostat and the system in it imposing the constraints
\begin{equation}\label{thermos}
    \lan\vf|(\pi_a(x)-p_a)=(\pi_a(x)-p_a)|\psi\ran=0.
\end{equation}
Here $x$ runs points of the thermostat and $p_a$ are fixed values of its intensive parameters. These constraints are preserved by the evolution and just say that the intensive variables of the thermostat do not fluctuate. The Hamiltonian corresponding to \eqref{hamilton func 1 field} is Hermitian and has a nonpositive spectrum. Hence any state of the system satisfying \eqref{thermos} tends to the ground state described by the probability density functional of the expected form
\begin{equation}
    \rho[\phi]\propto\exp\left[2\nu^{-1}\left(S_{sys}[\phi]-\int_{sys} dxp_a\phi^a(x)\right)\right],
\end{equation}
where the integral is taken over the system in the thermostat and $S_{sys}[\phi]$ is the thermodynamic entropy of this system. As before, the average entropy production $\dot{\hat{\Si}}$ relative to this state vanishes. For local $g^{ab}$ the density of fluctuational entropy production is zero in the thermostat, while it is of the form \eqref{entropy prod st micr} in the system. As a matter of fact, the matrix $g^{ab}(x,y)$ depends on the fields $\phi^a$. A generalization to this case is easily realized along lines \cite{TakWat,Grah,sd}.

\section{Concluding remarks}

Let us mention some possible modifications and generalizations of the formalism evolved here. Notwithstanding we distinguish extensive and intensive parameters, we did not actually use these properties. If the thermodynamic system possesses the ``gauge'' symmetry
\begin{equation}\label{extensiveness}
    x^i\partial_iS=S,
\end{equation}
i.e. $S(x)$ is a homogeneous function, a division into intensive $I$ and extensive $E$ variables is achieved by
\begin{equation}
    \de I:=\{x^iT_i,I\}\approx0,\qquad\de E:=\{x^iT_i,E\}\approx E.
\end{equation}
As it follows from \eqref{extensiveness} the Hessian is degenerate. Therefore we have to fix the extensive parameter $x^d$ in \eqref{1law} and work in the sector of remaining extensive variables and their conjugate. The fixed extensive parameter and its conjugate are expressed in terms of the independent ones by means of \eqref{extensiveness}. To put it in another way, we fix the ``gauge''. The division into extensive and intensive variables restricts admissible canonical transformations of the phase space of a thermodynamic system. Such a separation is preserved by arbitrary changes of intensive parameters and only linear transformations (with coefficients depending on the intensive variables) of extensive ones. In geometric terms, we realize the phase space as a Lagrangian fiber bundle \cite{Arn} with the base parameterized by intensive variables and the fiber represented by the configuration space.

Noteworthily, the thermodynamic (mean field) approach to phase transitions is naturally treated in the framework of symplectic geometry. A set of singular points of the Lagrangian surface, where the Hessian $\partial_{ij}S$ is degenerate, projected to the base of the Lagrangian bundle is called in optics and mechanics the caustic. In our case, the degeneracy condition bounds a region of extensive parameters, where the system is unstable, metastable states being regarded as stable. The caustics surround phase equilibrium curves (surfaces) on the base. When the system moves to the phase equilibrium curve and intersects the caustic, an additional local maximum in the entropy $\Si(x)=S(x)-p_ix^i$ of the whole system appears. This maximum corresponds to a new phase and becomes equal to the entropy maximum of the old phase on the equilibrium curve. It is interesting to apply developed methods of symplectic geometry to a study of topology and general forms of singularities \cite{Haken,Bucki,Gilm,PoSte,Thom,Jane} of Lagrangian surfaces, and, consequently, caustics and phase transition surfaces. The singularities of Lagrangian surfaces, which are stable with respect to small deformations respecting the Lagrangian bundle structure, are classified in \cite{Arn} up to ten dimensional phase spaces. The normal forms of Lagrangian surface singularities in phase spaces of higher dimensions have moduli.

Sometimes it is useful to define an entropy of a thermodynamic system not as the Legendre transform of $\Phi$, but as follows (see, e.g., \cite{KapGal,MagTer,Strat,Bali})
\begin{equation}\label{entropy stat}
    \tilde{S}(x)=\ln\left[\int dp\exp\left(\sum\limits_{i=1}^{d-1}p_ix^i-\Phi\right)\right],
\end{equation}
where contours of integration in complex planes are taken so that the integral converges. In a macroscopic limit, this definition coincides with the standard one, what is easy to see using the WKB-method. The extensive variables $x^i$ in \eqref{entropy stat} are not averaged over the ensemble as in \eqref{1law}. Rather, they describe one system in it. The intensive variables characterizing this system are taken, by definition, to be $\tilde{p}_i:=\partial_i\tilde{S}$. They slightly differ from the intensive parameters $p_i$ of the thermostat:
\begin{equation}
    \partial_i\tilde S=\partial_iS+\frac12\partial_{ijk}S\partial_p^{jk}\Phi+\ldots
\end{equation}
Assuming the Onsager principle \eqref{ons prpl} is fulfilled, the system with the Hamiltonian \eqref{hamilton func 1} and the entropy \eqref{entropy stat} decays to the state, where a detailed balance takes place. In this state, the
characteristic function \eqref{char func} is strictly proportional to $\ln Z$ up to an irrelevant additive constant.

An entropy of a thermostat is usually described by the term $-p_ix^i$ in the total entropy $\Si(x)$, since the intensive parameters of the thermostat are assumed to be nonfluctuating. If such fluctuations become relevant, they can be naturally taken into account in the formalism of stochastic deformation by introducing mixed states. These states are sums of pure states with some weights
\begin{equation}\label{mixed state}
    \hat\rho=\sum\limits_p\hat\rho_pe^{\s(p)}.
\end{equation}
The function $\s(p)$ is proportional to the entropy of the thermostat, while $-p_ix^i$ mentioned above is the entropy of interaction, by analogy with the interaction part of an action functional. The diagonal element of \eqref{mixed state} is proportional to the conditional probability $\rho_p(x)=\vf(x)\psi_p(x)$. Though mixed states are not exhausted by those ones.

The analogy with mechanics suggests also possible generalizations of a simple Hamiltonian model \eqref{hamilton func 1}. For example, it is interesting to consider Hamiltonian dynamics on a nonlinear symplectic space describing a thermodynamic system. Passing into the Darboux coordinates we see that a noncanonical symplectic structure results effectively in a changing of thermodynamic forces and the probability distribution law of fluctuations. Noncanonical symplectic structures appear naturally \cite{KuToHa,ZuMoRo,McLen,Rober,TreJar} in the case, when small deviations from the Gibbs distribution \eqref{stat sum} exist
\begin{equation}
    x^i=\partial^i_p\Phi+\xi^i(p),
\end{equation}
where $\xi^i$ is a small nongradient vector field. This equality defines a Lagrangian surface of a noncanonical symplectic structure with magnetic fields. For example \cite{TreJar}, if the density matrix of the system is proportional to $\exp[-p_i\hat{x}^i-\la(p_i\hat{x}^i)^2/2+O(\la^2)]$, where $\la$ is a small parameter, then in the leading orders
\begin{equation}
    x^i=\partial^i_p\Phi_0-\la[\partial^{ij}\Phi_0p_j-\partial^i_p\Phi_0\partial^j_p\Phi_0p_j+\frac12\partial^{ijk}_p\Phi_0p_jp_k-\partial^{ij}_p\Phi_0\partial^k_p\Phi_0p_jp_k+\partial^i_p\Phi_0(\partial^j_p\Phi_0p_j)^2 ]+O(\la^2),
\end{equation}
where $\Phi_0:=-\ln Z_0$ with $Z_0$ taken from \eqref{stat sum}. The last term in the  brackets is nongradient. Defining a thermodynamic entropy as the Legendre transform of the Massieu function $\Phi_0$, we arrive at the first law of thermodynamics with magnetic fields. The gradient part can be absorbed into the entropy by a gauge transformation. Another evident generalization of the model \eqref{hamilton func 1} consists in introducing non-Abelian gauge fields. We only touched the problem in Sec. \ref{fluct} and formulated the ``matter'' dynamics. The next step is to introduce the action functional for the  gauge fields. The gauge symmetries of gauge fields' action seem to be spontaneously broken, though a form of the action and a mechanism of the symmetry breaking are the subjects for further research.

Besides, we saw that quantum and stochastic mechanics differ from each other by a symmetry group only. Namely, the Heisenberg-Weyl algebra is a Lie algebra with commutation relations
\begin{equation}
    [e_1,e_2]=e_3,\qquad[e_1,e_3]=[e_2,e_3]=0,
\end{equation}
where $e_1$, $e_2$ and $e_3$ are its generators and, for brevity, we consider a two-dimensional phase-space. This algebra includes an Abelian ideal spanned on $e_3$. A general form of the Lie group corresponding to this ideal is $U^n(1)\times SO^k(1,1)$. In the two simplest cases $n=1$, $k=0$ and $n=0$, $k=1$ we obtain quantum and stochastic mechanics respectively. So, one can speculate about their unification by introducing a larger group containing the subgroups $U(1)$ and $SO(1,1)$.


\begin{acknowledgments}

I am grateful to A.V. Shapovalov and A.A. Sharapov for reading a draft of this paper, discussions and useful suggestions. This work was supported by the RFBR Grant No. 09-02-00723 and Support of Russian Scientific Schools Grant No. SS-871.2008.2.

\end{acknowledgments}

\end{document}